\documentclass[twocolumn,twoside,prd,floatfix,letterpaper]{revtex4}

\usepackage{ifpdf}
\ifpdf 
  \usepackage[pdftex]{graphicx} 
\else 
  \usepackage{graphicx} 
\fi
\usepackage{amsmath}
\usepackage{amssymb}
\usepackage{fancyhdr}
\usepackage{journals}
\usepackage{color}
\usepackage{rotating}
\usepackage[colorlinks,hyperindex]{hyperref}

\def \informationgain {{\ensuremath{\text{information gain} (x_i)}}}

\def \noty {{\ensuremath{y\!\!\!\slash}}}
\def \yMary {{\ensuremath{y_{\text{Mary}}}}}
\def \yKathy {{\ensuremath{y_{\text{Kathy}}}}}
\def \ySarah {{\ensuremath{y_{\text{Sarah}}}}}
\def \logten {{\ensuremath{\text{log}_{10}}}}
\def \onepercent {{1\%}}
\def \expectedPayoff {{\ensuremath{\langle \text{payoff} \rangle}}}
\newcommand \bknot[1] {{\ensuremath{\not{#1}}}}

\begin{document}

\title{Capitalist Science}
\author{Bruce Knuteson}
\homepage{http://bruceknuteson.com/}
\email{knuteson@mit.edu}
\noaffiliation


\begin{abstract}
The economic structure of basic science is currently socialist, funded by the public at large through taxes for the benefit of the public at large.  This socialist system should be augmented by a capitalist system, in which basic science is also funded by private investors who reap financial benefit from the sale of subsequent technologies based on the knowledge obtained from the research funded by their investments.  A capitalist system will provide benefits extending from the broad target audience of this paper --- which includes politicians, financiers, economists, and scientists in all fields --- to the average taxpayer and consumer.  Capitalist science will better align the incentives of scientists with taxpayer interests, channel more money into basic science, lower your taxes, and generally improve the quality of your life.
\end{abstract}

\maketitle
\tableofcontents

\section{Socialist science}
\label{sec:SocialistScience}

Basic science is currently a socialist enterprise, funded by the taxpayer, with the results available to society at large.  Government agencies centrally plan the bulk of funding decisions, with peer review providing input to committees that assist these agencies in selecting projects for funding.  Nobody owns the knowledge that is produced, which is available at no cost to anyone capable of reading the resulting publications.  From each according to his income, to each according to whether he can read~\footnote{Gender equality is attempted by using the male singular personal pronoun in the main text, and female scientists in the example in Appendix~\ref{sec:Example}.}.

This socialist science leads to a classic tragedy of the commons~\footnote{The maintenance of a shared resource (such as a communal area, or ``commons'') tends to be neglected.  Any individual who puts effort into maintaining the commons shoulders the cost of this effort, while the benefits are widely distributed among the population.  Responding rationally to incentives, the typical individual will focus his efforts on activities that benefit himself more than maintaining such a shared resource.  All individuals thereby tragically suffer from individually rational actions.  Game theory is a field that studies this and similar problems.}.  Since the benefits of scientific advances are consumed by everyone, no individual has a strong financial incentive to invest in basic research.  Elementary game theory suggests that basic science is underfunded as a direct result of these conditions.  More simply stated, socialist science leads to actions by individuals responding rationally to their incentives that may not be in the best interest of society at large.  To the extent such a problem exists, the cure is not to fault the individuals, but rather to change the incentives.

It is commonly believed that governments are the institutions with the longest term perspective, and that therefore governments should be responsible for funding basic science.  This common perception belies the fact that the interests of the individual decision makers in government are actually much shorter in horizon than the interests of a patient investor.  A politician focused on winning his seat in the next election is unlikely to fight hard to fund some basic science that will pay off globally over the next decade.  The politician who focuses on the immediate needs of his constituents is behaving perfectly rationally in response to his incentives, so faulting him is not helpful.

Decisions about what science to fund are made by human beings, and they are therefore obviously affected at some level by the personal interests of the decision makers.  Each scientist can be expected to have a vested interest in his own research agenda, and to react more favorably to proposals that further that agenda than to those that suggest a different direction.  Few scientists are apt to welcome funding to which they have grown accustomed being redirected to an area in which they see themselves having no particular competitive advantage, since the switching costs inherent in changing research direction can be significant.  If this approach results in inefficient science funding decisions, the blame should not be placed on the scientists involved in those decisions, who are behaving perfectly rationally in response to their incentives.  Nor should scientists, who as a group expend considerable effort attempting to ensure a fair and thoughtful process of review, be accused of harboring an unwillingness to do the right thing.

Younger scientists who may not yet be involved in funding decisions perform their research in the context of an incentive structure that varies somewhat from field to field.  Across most fields, number of publications and number of citations are two often used measures of a scientist's contribution to his field.  Young scientists therefore naturally focus on producing publications that are apt to be highly cited, and those that do this most successfully are promoted and have the opportunity to become old scientists.  The use of measures like number of publications and number of citations may seem reasonable, appearing to capture both the quantity and impact of a scientist's work.  Unless the goal of the average American taxpayer is to maximize the total number of citations, there is a serious disconnect between the goals of the young scientist and the goals of those ultimately funding his research.  The young scientist is not culpable since he is behaving perfectly rationally in the context of his environment, and he can hardly be blamed for the evolved misalignment of his incentives and those of the taxpayer.  

Because socialist science ensures underfunded science through a tragedy of the commons, many young scientists will grow into something other than old scientists.  The young scientist who seeks an alternative occupation in response to price signals is very probably acting rationally in response to his incentives.  In particular, the fact that his career change may slightly harm society at large, which could benefit more from the basic research advances he would make than from whatever he will do in his new and more highly compensated occupation, is not really his problem.

The underfunding of science results in some useful science not getting done.  As one of many examples, the climate change problem currently being faced by the world is a large one, and one that could benefit from additional basic science research related to the problem.  Incentives for individual scientists to perform basic research relevant to the problem are not of the same magnitude as the problem.      

Given the value most scientists place on measurement, current technology for quantitatively tracking scientific progress is primitive.  Very few papers explicitly quantify how beliefs in the clearly and fully articulated hypotheses considered should be updated in light of the work presented therein.  The resulting ambiguity severely hampers attempts to draw broader quantitative conclusions from multiple related papers.  These observations come as no surprise, since the current funding process provides no particular incentive for authors to quantify results in a manner amenable to further quantitative inference.

For those unwilling to measure scientific progress by the total number of citations, tracking scientific progress requires tracking how beliefs in various proposed hypotheses change.  For the public at large, changing beliefs in proposed hypotheses are currently tracked in occasional news stories, with particularly interesting dramas summarized long after their time in the odd history of science book.  Changes of beliefs in proposed hypotheses are not available in real time, on the web.  Imagine if financial information were not available in real time, on the web, and the only information anyone received regarding changes of beliefs in hypotheses related to the modern economy were from occasional news stories and the odd popular book.  Although such limited information availability might be suitable for somebody following along at home, it is wholly unsuitable for the active practitioner.  Socialist science, which lacks incentives for such real time tracking of beliefs in proposed hypotheses, is missing a market consensus.  The extent to which an absence of quantitative, real time belief tracking has been slowing scientific progress will not be widely appreciated until it is commonplace~\footnote{With a hypothesis database, human driven climate change might have received widespread acceptance a decade earlier.  In general, forcing all discussion participants to put serious money where their mouths are significantly changes many discussions.}.

The problems with socialist science, although not sufficiently disastrous to prevent some good science from getting done, may motivate brief consideration of an alternative.  The purpose of this paper is to propose that the current socialist science system be augmented with capitalist science.  The proposal is described in Sec.~\ref{sec:CapitalistScience}, with Sec.~\ref{sec:Summary} summarizing.  Appendix~\ref{sec:Criticisms} provides some additional commentary, noting potential criticisms and providing brief rejoinders.  For those interested in more technical details, Appendix~\ref{sec:Example} provides an example of how this might look in practice, and Appendix~\ref{sec:PreviousWork} makes a connection to the related ideas of Refs.~\cite{ExperimentalScientificMerit:Knuteson:2007tb,TheoreticalScientificMerit:Knuteson}.  The proposal is a minor extension of ideas that are already mainstream:  that capitalism has certain advantages over socialism, that basic science tends to spur subsequent technological innovation and improvement of standard of living, and that individual incentives matter.

\section{Capitalist science}
\label{sec:CapitalistScience}

The proposal advanced in this article is nonetheless sufficiently different from current practice that it may take some time before the underlying logic is fully appreciated.  The reader will be forgiven if the overview provided in this section strikes him as fanciful until after reading Appendices~\ref{sec:Criticisms} and~\ref{sec:Example}, which describe in more detail how this will work in practice.  It may be useful to think of the proposal (very loosely) as a cross between financial markets and the patent system, both of which manage to function roughly as intended, but each of which would be laughable if introduced from scratch to a new audience in a three page paper.  

Augmenting socialist science with capitalist science will require an act of Congress.  New legislation needs to be passed that applies a small tax on companies' revenue, starting some time (say ten years) in the future.  This tax is distributed in full to those responsible for the knowledge making the products and services of these companies possible, as detailed in Appendix~\ref{sec:Example}.  Those who bristle at the word ``tax'' should note that this proposal represents a direct redistribution rather than an increase in funds over which the government has spending discretion.  As discussed in Appendix~\ref{sec:Criticisms}, this proposal will actually decrease taxes.  Rather than asking current taxpayers to fund science they do not understand for the benefit of who knows who at some undetermined point in the future, the people in the future who actually benefit from the knowledge gained will pay the people who fund today's research.

Using round numbers for the purpose of illustration, suppose the tax rate Congress decides to apply to companies' revenues starting in ten years is 1\%.  If the gross domestic product of the United States ten years from now is \$20T, then ten years from now a total of \$200B will be paid to people who have invested in scientific research in the interim.  Applying an appropriate discount for the passage of time, there exist reasonable accounting choices for which this tax revenue roughly approximates the current funding of socialist science.

The \$1 or so of tax on each \$100 of company revenue is divided among those who have contributed to producing the knowledge necessary for their products and services to work as intended, in a manner described in Appendix~\ref{sec:Example}.  The division rewards studies that have significantly increased beliefs in hypotheses whose correctness is necessary for their products or services to work as designed.  

Given the amount of money involved, new and existing companies start focusing on basic science shortly after the legislation is passed.  Basic science suddenly becomes entrepreneurial, a quality that has been lost in some more mature scientific fields.  A person with a clever idea can actually get paid for generating useful knowledge about how the world works.  Funds raise money from investors to sponsor the development of these clever ideas.  Some of the resulting research enterprises grow over time and become publicly traded companies, with revenue coming from the people buying products whose development has relied on the improved understanding of nature resulting from the company's research.  The investors who fund tomorrow's scientific breakthroughs receive improved risk adjusted returns through the additional diversification this investment alternative provides.  Portfolio managers with science backgrounds have a financial incentive to identify the most promising research.  Many of the best scientists start funding their research from private money rather than from government grants, since research done with private money has a financial upside not present in government funded research, which is ineligible for a cut of the tax on companies' revenue.  With fewer scientists continuing to rely on government funding, the fraction of the federal budget devoted to supporting socialist science gradually decreases, and portions adequately replaced by capitalist science are eventually largely phased out.

Professional forty-year-olds with another forty years ahead of them, representing a group with significant aggregate investment capacity, have a risk tolerance commensurate with the risk of basic research.  The size of this investment pool is an important determinant in how quickly capitalist science supplants socialist science, since these are the investors who are funding today's capitalist science.  Investments in companies (both public and private) that perform basic science research become a standard part of the average investor's portfolio, with a portfolio weight that decreases with investor age.  The same secondary market that allows an investor to cash out of his investment in IBM before realizing all future cash flows allows an investor to cash out of his investment in a basic science company before realizing all future cash flows, with the usual constraint that he can only sell his investment at a price somebody else is willing to pay.  Funds are developed to provide individual investors diversified exposure to a range of scientific research, in the same way that some mutual funds today currently provide an easy way for investors to obtain diversified exposure to a range of stocks of companies with traditional revenue streams.

A database of beliefs in proposed hypotheses is used to track the changes in beliefs on which payouts are based, as described in Appendix~\ref{sec:Example}.  A study testing some hypothesis against its negative may motivate a database update that quantifies the authors' view of how the study performed should affect our belief in the correctness of that hypothesis.  Scholarship is no longer published in academic journals.  Publication takes the form of a database entry, consisting of the authors' specification of one or more numbers between zero and unity reflecting updated beliefs in one or more hypotheses, together with a non-peer-reviewed paper providing the details of the study.  A single database update may simultaneously change beliefs in multiple hypotheses.  New hypotheses are entered through a database update consisting of a complete articulation of the hypothesis and an initial belief that the hypothesis is correct.  The current state of the database, all updates to the database, and the complete history of the database are all fully accessible to everyone at no cost.

Each company identifies those hypotheses in the database whose veracity (or falsity) is necessary for their products or services to work as intended.  When finalized, money is moved from the company's account to and among the accounts of those responsible for relevant database entries.  Each author receives a fraction of the revenue proportional to the information gain he has provided in favor of nature working in such a way that each product or service functions as intended, as specified in more detail in the example in Appendix~\ref{sec:Example}.  Claims on future revenue expire fifty years after publication.  Authors who turn out to be wrong owe an amount proportional to the negative information gain they have provided.  All payments are tracked and fully publicly disclosed on the web, with disputes handled by an administrative agency.

The resulting market is big enough to attract arbitrageurs who systematically look for beliefs that are likely to be incorrect, and who make their money by updating the database with more accurate beliefs.  Since participants can lose money by being wrong, adequate insurance is required for any participant desiring to assume significant exposure.  The database itself needs to be robust but not particularly complicated, and is run by a small but competent group within the Office of Science and Technology Policy.

\section{Summary}
\label{sec:Summary}

The economic structure of basic science is currently socialist, funded by the public at large through taxes for the benefit of the public at large.  This socialist science should be augmented by capitalist science, in which basic science is also funded by private investors who reap financial benefit from the sale of subsequent technologies based on the knowledge obtained from the research funded by their investments.    

A short article intended for a broad readership clearly cannot hope to address all issues that could be raised of a proposal of this scale.  Appendix~\ref{sec:Criticisms} provides brief, nontechnical responses to some anticipated questions.  Technical readers interested in perusing a worked example are encouraged to consult Appendix~\ref{sec:Example}, and those already familiar with the arguments of Refs~\cite{ExperimentalScientificMerit:Knuteson:2007tb,TheoreticalScientificMerit:Knuteson} may find helpful the tie to previous work made in Appendix~\ref{sec:PreviousWork}.  

The heart of the idea proposed in this article is a restructuring of incentives that better aligns the interests of scientists and those funding their research.  Expected direct effects include more money being channeled into basic science despite a reduction in government spending, better use of this money from the viewpoint of the taxpayer, and a resultant improved quality of life when you retire.

\acknowledgments
Bill Ashmanskas, Susan Burkhardt, Tom Ferbel, Bill Foster, Ted Liu, Stephen Mrenna, Allen Poteshman, and Huafeng Xu provided helpful comments on initial drafts.  Equations are absent from the main text at the gentle suggestion of the author's parents.  

\appendix

\section{Criticisms and rejoinders}
\label{sec:Criticisms}

A few anticipated miscellaneous criticisms and brief responses are provided in a bit of give and take below.

{\em{Won't this stifle the openness of basic science?}}  No.  This proposal will actually increase the openness of basic science by providing a financial incentive for publishing potentially useful results.

{\em{Are you raising taxes?}}  No.  The proposed tax on companies' revenue starting on some future date that will pay today's investors in basic science does not change the pool of money over which the government has spending discretion, since the full amount of the tax goes directly and formulaically to private investors.  To the extent government funded science is successfully replaced by capitalist science, the public's tax burden is actually decreased.

{\em{Why 1\%?}}  A tax rate of \onepercent\ is used as a round number for the purpose of this paper.  Suppose the tax is set to start in ten years, or perhaps is set to phase in gradually from five to fifteen years hence, starting at 0.1\% in 2016 and increasing by 0.1\% per year to 1.0\% in 2025.  The net present value of the tax revenue per year ten years out approximates the government's annual science funding today.  The amount of private money invested in basic science will be proportional to whatever rate is set.  A rate on the order of 1\% is probably large enough to incentivize basic research without being so large that it disincentivizes commercialization of what is learned.  It also happens to be roughly the rate skimmed by the financial sector on every transaction paid for with plastic, a service that arguably provides far less social value.

{\em{How can you possibly determine the expected future tax revenue two decades from now that might come from some basic science research done today?}}  In much the same way you value the worth of owning a share of IBM:  with difficulty.  Of course, this difficulty in valuation is precisely why you expect (in aggregate) to receive a greater return on your (riskier) investment in basic science than on your (less risky) investment in a blue chip corporate bond.

{\em{How does this differ from patents?}}  Speaking loosely, patents are granted for inventions that do something you might want to do.  A patent gives the patent holder the right to exclude others from utilizing the invention.  In contrast, capitalist science gives people who increase our knowledge of how nature works a claim on a fixed fraction of future cash flows from the sale of those products and services that rely on this knowledge.

{\em{What keeps authors from overstating their results?}} Authors have to pay if they are wrong, as described in Appendix~\ref{sec:Example}.  Assuming a set of hypotheses with equal expected revenue streams, a simple calculation shows that self-interested, risk neutral scientists will want to publish database updates equal to their actual beliefs to maximize their expected payoff.  Self-interested, risk averse scientists will systematically understate their results, publishing ``conservative'' updates that deviate less from existing database values than their analyses imply.

{\em{This hypothesis database actually seems pretty cool.  Can I have it without changing science funding?}}  No.  Beliefs in the hypothesis database are valuable precisely because there is money on the line.  Although this article leads with a capitalist approach to science funding and notes, seemingly incidentally, the necessity of a hypothesis database, it is equally valid to lead with the goal of a hypothesis database and note, seemingly incidentally, the necessity of a capitalist approach to science funding.

{\em{What keeps companies from misidentifying hypotheses that need to be correct for their products and services to function as intended?}}  All companies will lose \onepercent\ of revenue, with the bulk of this loss presumably passed along to the consumer.  Misallocating the \onepercent\ is apt to trigger a dispute, the anticipated cost of which incentivizes the company to allocate correctly.

{\em{Since the paper submitted by the authors as part of a database update is not subject to peer review, and since all that seems to matter is the changes the authors make to beliefs, what motivates authors to write a high quality paper explaining what they have done?}}  Authors who feel disadvantaged by some falsehood or obscurity in a paper submitted with another author's database entry are able to initiate an administrative proceeding against that author for revenues they believe they lost due to that falsehood or obscurity.  This recourse gives authors a financial incentive to produce clear, detailed papers to accompany database updates.

{\em{You seem to be relying on an administrative agency for all of the messy decisions.  Aren't you just passing the buck to a bunch of lawyers?}}  No.  In most cases the system will work automatically, as advertised, and without lawyers.  In a few cases, just as with patents, disputes will need to be resolved by humans with appropriate expertise.

{\em{Why should I believe the government will follow through and actually impose a future 1\% tax, even if it promises to do so now?}}  The government's enacting of capitalist science will create many new jobs, for nonscientists as well as scientists, as investment funds are created to channel money into basic science in anticipation of the promised tax revenue.  By failing to impose the \onepercent\ tax at some point in the future after promising to do so, the government would obliterate the jobs created in anticipation of this tax revenue.  For better or worse, the government has historically been quite responsive to political pressure arising from concentrated job losses.

{\em{Won't the tax, which can only be applied to domestic companies, give foreign companies an advantage?}}  Adopting a policy whereby authors of database entries from other countries get their share if and only if company revenues in that country are subject to the same tax ensures that the exchange of money is contained within the set of participating countries.  The adoption of capitalist science by other countries, as with patent protection in much of the developed world, carries certain advantages.  

{\em{Won't this proposal lead to more shortsighted research?}}  This proposal may in many ways actually lead to decisions with greater foresight.  As a specific example, with the exception of Refs.~\cite{QuaeroD0:Abazov:2001ny2, QuaeroH1:Caron:2006fg2}, little to no attempt is made to retain data from expensive particle physics experiments, since it is not really in the interest of any individual physicist to go to the trouble.  Capitalist science provides a direct financial incentive to retain data from an expensive study in a manner that lends itself to the quick testing of additional hypotheses in the future.

{\em{Do you seriously think Congress might pass this?}}  In addition to its intrinsic merit, this proposal has elements that should appeal to both political parties.  The first half of the bipartisan, two word title of this article is stereotypically associated with Republicans, and the second with Democrats.

{\em{Any policy change has winners and losers.  Who wins from this proposal, and who loses?}}  
The primary motivation for this proposal is the expectation that society will be better off thirty years from now if it is adopted than if it is not.  More quality basic science will get done, which will lead to increased knowledge, improved products and services, and increased quality of life.  The average taxpayer wins because he gets more by paying an extra 1\% on things that might be otherwise unavailable or unaffordable than by paying taxes for basic science that may never result in something he will use.  Investors win with greater risk adjusted returns due to the additional diversification that investment in basic science provides.  Clever, innovative scientists win because new breakthroughs are accompanied by significant financial upside.  The only people who lose are scientists receiving taxpayer funding for doing research that lacks the promise of producing interesting technologies within a typical young person's lifetime.  For understandable, self-interested reasons, this group is likely to raise the most vociferous objections to this proposal.

\section{Example}
\label{sec:Example}

A simplified example illustrates how the proposal outlined above would work in practice.  This appendix (and the one following) are somewhat technical, and are targeted at a much narrower readership than the main text of this article.   

A hypothesis database is used to track changes in beliefs of proposed hypotheses.  Although updates to the hypothesis database occur in continuous time, for bookkeeping purposes the $t^{\rm{th}}$ database update can be considered to occur at time $t$.  A generic analysis or experiment $x$ affects our belief in some hypothesis $y$~\footnote{With slight abuse of notation, $x$ may represent either the analysis itself or the outcome of the analysis.}.  An analysis or experiment $x_t$ motivates an update to the database at time $t$.  A subscript on $y$ labels a specific hypothesis.  The probability (likelihood) of seeing the outcome $x$ assuming the correctness of the hypothesis $y$ is $p(x|y)$.  The probability of obtaining the result $x_t$ assuming the correctness of the hypothesis $y$ and given the knowledge of database entries up to and including time $t-1$ is $p(x_t|y, t-1)$, where a single number ($t-1$) encapsulates all information available in the database up to and including time $t-1$.    In general, any analysis or experiment will test competing hypotheses, such as a hypothesis $y$ and its negative $\noty$.  Such an analysis $x_t$ will then lead to an update to the database $p(y|t)$, a number between zero and unity reflecting the belief that $y$ is correct at time $t$, by applying standard Bayesian logic to information available in the database at time $t-1$.

A new hypothesis is entered through a database update consisting of a complete articulation of the hypothesis and a value for the prior, which is an initial belief that the hypothesis is correct.  This update is public, as is the statement of the hypothesis and the proposed value for the prior.  To ensure a fair initial belief, such an update starts an auction period lasting two weeks.  During this time other participants have an opportunity to submit different values for the prior.  No information submitted during the auction becomes public until the auction period ends, at which time all submitted information becomes public.  Upon the conclusion of the auction, the prior on the hypothesis is set equal to the geometric mean of the submitted values.  Every participant submitting a prior has economic exposure to the value he submits.  The rationale for this auction should become clear through the example provided in this appendix.

The example narrated below is summarized as entries into a toy hypothesis database in Table~\ref{tbl:desresExampleDatabaseEntries}, with each numbered paragraph $t$ corresponding to one database entry.  These entries give rise to a database whose cumulative state at time $t$ is summarized in Table~\ref{tbl:desresExampleDatabaseState}.

\begin{table}
\caption{Entries into the toy database corresponding to the example narrated in the text.  Each numbered row represents the database entry resulting from the correspondingly numbered paragraph in the text.  The entry's author is given in the second column, and the entry itself is provided in the third column.  These entries produce a database whose state at time $t$ is shown in Table~\ref{tbl:desresExampleDatabaseState}.  $^\dag$The first database entry of a hypothesis is determined by an auction, in which values for priors are solicited and a geometric mean of these values is entered in the database as the hypothesis prior.\\}
\label{tbl:desresExampleDatabaseEntries}
\begin{tabular*}{0.4\textwidth}{@{\extracolsep{\fill}} c | l | l c l}
$t$& Author   &  Entry        &   &                     \\ \hline
1  & \dag     &  $p(\yMary)$  & = & {\tt{0.000206}}     \\
2  & Dawn     &  $p(\yMary)$  & = & {\tt{0.03}}         \\
3  & Mary     &  $p(\yMary)$  & = & {\tt{0.9999999}}    \\
4  & \dag     &  $p(\yKathy)$ & = & {\tt{0.00000728}}   \\
5  & Kathy    &  $p(\yKathy)$ & = & {\tt{0.9999999}}    \\
6  & Anne     &  $p(\yMary)$  & = & {\tt{0.9999}}       \\
7  & Mary     &  $p(\yMary)$  & = & {\tt{0.9999999999}} \\
8  & \dag     &  $p(\ySarah)$ & = & {\tt{0.0299}}       \\
9  & Sarah    &  $p(\ySarah)$ & = & {\tt{0.999}}        \\
\end{tabular*}
\end{table}

\begin{table}
\caption{State of the toy database corresponding to the example narrated in the text, providing beliefs in the hypotheses \yMary, \yKathy, and \ySarah\ at each time $t$.  The state of the database at time $t$ reflects the database entries in Table~\ref{tbl:desresExampleDatabaseEntries} up to and including time $t$.  New database entries are marked with an asterisk.\\}
\label{tbl:desresExampleDatabaseState}
\begin{tabular*}{0.4\textwidth}{@{\extracolsep{\fill}} c | l l l}
$t$&   {$p(\yMary)$} & {$p(\yKathy)$} & {$p(\ySarah)$} \\ \hline
1  &  {\tt{0.000206}}*    &                    &                \\
2  &  {\tt{0.03}}*        &                    &                \\
3  &  {\tt{0.9999999}}*   &                    &                \\
4  &  {\tt{0.9999999}}    & {\tt{0.00000728}}* &                \\
5  &  {\tt{0.9999999}}    & {\tt{0.9999999}}*  &                \\
6  &  {\tt{0.9999}}*      & {\tt{0.9999999}}   &                \\
7  &  {\tt{0.9999999999}}*& {\tt{0.9999999}}   &                \\
8  &  {\tt{0.9999999999}} & {\tt{0.9999999}}   &  {\tt{0.0299}}*\\
9  &  {\tt{0.9999999999}} & {\tt{0.9999999}}   &  {\tt{0.999}}* \\
\end{tabular*}
\end{table}

\begin{enumerate}
\item Mary suspects various cancers all have a common feature.  She realizes she can test her hypothesis \yMary\ conclusively at low marginal cost given her company's existing infrastructure.  Mary suspects her field would estimate the probability of \yMary\ being correct at $10^{-4}$, but she has obtained preliminary data that cause her to estimate this probability at $10^{-3}$.  Mary convinces her company to enter a prior of $2 \times 10^{-4}$ into the hypothesis database.  It is in Mary's company's interest to enter Mary's hypothesis and a value for the prior into the database, since otherwise they will not make any money publishing the results of the analysis Mary intends to perform.  During the auction period over the next two weeks, additional priors of $3\times10^{-4}$, $3\times10^{-5}$, and $10^{-3}$ are entered.  When the auction has closed at time $t=1$, $p(\yMary|1)$ is set at $2.06 \times 10^{-4}$, the geometric mean of the four priors.

\item Mary then performs her test over the next few months.  Before she is ready to publish, her former college roommate Dawn, working at a competing company, performs a quick and dirty analysis $x_2$ and finds $\logten{\frac{ p(x_2 | \yMary, 1) }{ p(x_2 | \bknot{\yMary}, 1)}} = +2.13$.  Applying the usual rules of Bayesian inference~\footnote{Applying the usual rules for Bayesian inference, the database update $p(y|t)$ is obtained by solving 
$\frac{p(y|t)}{p(\noty|t)} = \frac{p(x_t|y, t-1)}{p(x_t|\noty, t-1)}\frac{p(y|t-1)}{p(\noty|t-1)}$ for $p(y|t)$, subject to the logical constraint $p(y|t') + p(\noty|t') = 1$ for all times $t'$.  On the right hand side of this equation, the first fraction is the likelihood ratio obtained from the analysis $x_t$, and the second fraction is taken from the database.}
, she publishes $p(\yMary | 2) = 3 \times 10^{-2}$ to the database.

\item After significant discussion within Mary's company regarding her analysis $x_3$, her company concludes $\logten{\frac{ p(x_3 | \yMary, 2) }{ p(x_3 | \bknot{\yMary}, 2)}} = +9.2$, and publishes $p(\yMary | 3) = 1 - 10^{-7}$ to the database.

\item Shortly thereafter, Kathy suspects that this feature shared by various cancers may be altered when in solution with a substance possessing a particular characteristic~\footnote{Details to be filled in in the presence of improved financial incentives.}.  Kathy suspects her field would estimate the probability of \yKathy\ being correct at $10^{-6}$, but she has an argument that causes her to put this probability at $10^{-2}$.  Kathy convinces her company to enter a prior of $1.5\times10^{-5}$ into the hypothesis database.  During the auction period over the next two weeks, additional priors of $4\times10^{-4}$, $1.7\times10^{-5}$, $10^{-6}$, and $2\times10^{-7}$ are entered.  When the auction has closed, $p(\yKathy|4)$ is set at $7.28 \times 10^{-6}$, the geometric mean of the five priors.

\item Kathy then performs her analysis $x_5$ over the next year, eventually finding $\logten{\frac{ p(x_5 | \yKathy, 4) }{ p(x_5 | \bknot{\yKathy}, 4)}} = +12.6$, and convinces her company to publish $p(\yKathy | 5) = 1 - 10^{-7}$ to the database.

\item Anne performs a study $x_6$, spots what she thinks is an error in Mary's analysis, argues $\logten{\frac{ p(x_6 | \yMary, 5) }{ p(x_6 | \bknot{\yMary}, 5)}} = -3.2$, and publishes $p(\yMary | 6) = 1 - 10^{-4}$ to the database.

\item Mary defends and extends her previous analysis in a study $x_7$, argues $\logten{\frac{ p(x_7 | \yMary,6) }{ p(x_7 | \bknot{\yMary},6)}} = +7.5$, and convinces her company to publish $p(\yMary | 7) = 1 - 10^{-10}$ to the database.

\item Soon thereafter, Sarah realizes it may be possible to prevent cancer cells from replicating with a treatment that uses a particular substance with the characteristic noted by Kathy to attack the common feature noted by Mary.  Sarah convinces her company to publish a prior of $2 \times 10^{-2}$ on her hypothesis \ySarah, which seems somewhat plausible given the current state of the field.  During the subsequent auction period, priors of $3\times10^{-2}$, $10^{-2}$, $8\times10^{-2}$, and $5\times10^{-2}$ are entered.  When the auction has closed, $p(\ySarah|8)$ is set at $2.99\times10^{-2}$, the geometric mean of the five priors.

\item Sarah then quickly does an analysis $x_9$, finds $\logten{\frac{ p(x_9 | \ySarah, 8) }{ p(x_9 | \bknot{\ySarah}, 8)}} = +5.1$, and convinces her company to publish $p(\ySarah | 9) = 1 - 10^{-3}$ to the database.

\end{enumerate}

Six years later, after clinical tests, a treatment is available that cures several forms of cancer in 90\% of patients with tolerable side effects.  The product begins to save millions of lives per year, and generates a revenue of \$4.2B in 2020 for companies operating in or engaging in trade with the United States. 

The companies selling this product are required by the Capitalist Science Act of 2012 to identify the hypotheses in the database (or their negatives) that are necessary for the product to work as intended.  By 2020, the database contains millions of hypotheses.  Of the hypotheses in the database in 2020, many have a posterior (current belief) sufficiently close to the prior (set by the initial auction for that hypothesis) that the money made by the authors who provided evidence for the hypothesis is negligible~\footnote{Both the prior (circa 2012) and posterior on Newton's Law of Gravity (with appropriate caveats) will be very close to unity, for example, and the money made by the authors of database entries adjusting the belief in Newton's Law of Gravity can be neglected in this example.}.  Among the hypotheses in the database in 2020 for which the evidence provided is non-negligible, only \yMary, \yKathy, and \ySarah\ are identified as being necessary for the cancer cure to function as intended.  Authors who provided evidence in favor of two other related hypotheses initiate an administrative proceeding, arguing that these two related hypotheses also need to be true for the product to function as intended, but the administrative agency charged with handling such disputes concludes that the product could plausibly function as intended even if these two related hypotheses are not true.

\begin{table}
\caption{Evidence provided in each database entry in the example narrated in the text.  The evidence provided in favor of hypothesis $y$ by database entry $t$ is $\logten{\left(p(y|t)/p(y|t-1)\right)}$.  Payouts, determined by dividing this evidence by the total information gain calculated in Eq.~\ref{eqn:totalInformationGain} and multiplying by 1\% of the total revenue, are shown (in \$M) in the rightmost column.  $^\dag$The first database entry of a hypothesis is an auction, in which values for priors are solicited and a geometric mean of these values is entered as the hypothesis prior.  Participants in these auctions other than this example's main cast of characters are simply labeled ``Bidder''.  Each auction is a zero sum game, with a net payout of \$0.\\}
\label{tbl:desresExampleEvidence}
\begin{tabular*}{0.475\textwidth}{@{\extracolsep{\fill}} c | l | r c | r}
$t$& Author   &  Evidence                 & Hypothesis  &  \multicolumn{1}{c}{Payout (\$M)}\\ \hline
1  & \dag     &                           & \yMary                                \\
   & Mary     &  {\tt{\verb|- 0.01   |}}  &             &  {\tt{\verb|- 0.05  |}} \\
   &Bidder \#1&  {\tt{\verb|+ 0.2    |}}  &             &  {\tt{\verb|+ 0.7   |}} \\
   &Bidder \#2&  {\tt{\verb|- 0.8    |}}  &             &  {\tt{\verb|- 3.4   |}} \\
   &Bidder \#3&  {\tt{\verb|+ 0.7    |}}  &             &  {\tt{\verb|+ 2.8   |}} \\
2  & Dawn     &  {\tt{\verb|+ 2.2    |}}  & \yMary      &  {\tt{\verb|+ 8.8   |}} \\
3  & Mary     &  {\tt{\verb|+ 1.5    |}}  & \yMary      &  {\tt{\verb|+ 6.2   |}} \\
   &          &                           &             &                         \\
4  & \dag     &                           & \yKathy     &                         \\
   & Kathy    &  {\tt{\verb|+ 0.3    |}}  &             &  {\tt{\verb|+ 1.3   |}} \\
   &Bidder \#1&  {\tt{\verb|+ 1.7    |}}  &             &  {\tt{\verb|+ 7.1   |}} \\
   &Bidder \#2&  {\tt{\verb|+ 0.4    |}}  &             &  {\tt{\verb|+ 1.5   |}} \\
   &Bidder \#3&  {\tt{\verb|- 0.9    |}}  &             &  {\tt{\verb|- 3.5   |}} \\
   &Bidder \#4&  {\tt{\verb|- 1.6    |}}  &             &  {\tt{\verb|- 6.3   |}} \\
5  & Kathy    &  {\tt{\verb|+ 5.1    |}}  & \yKathy     &  {\tt{\verb|+20.9   |}} \\
   &          &                           &             &                         \\
6  & Anne     &  {\tt{\verb|- 0.00004|}}  & \yMary      &  {\tt{\verb|- 0.0002|}} \\
7  & Mary     &  {\tt{\verb|+ 0.00004|}}  & \yMary      &  {\tt{\verb|+ 0.0002|}} \\
   &          &                           &             &                         \\
8  & \dag     &                           & \ySarah     &                         \\
   & Sarah    &  {\tt{\verb|- 0.2    |}}  &             &  {\tt{\verb|- 0.7   |}} \\
   &Bidder \#1&  {\tt{\verb|+ 0.001  |}}  &             &  {\tt{\verb|+ 0.004 |}} \\
   &Bidder \#2&  {\tt{\verb|- 0.5    |}}  &             &  {\tt{\verb|- 1.9   |}} \\
   &Bidder \#3&  {\tt{\verb|+ 0.4    |}}  &             &  {\tt{\verb|+ 1.7   |}} \\
   &Bidder \#4&  {\tt{\verb|+ 0.2    |}}  &             &  {\tt{\verb|+ 0.9   |}} \\
9  & Sarah    &  {\tt{\verb|+ 1.5    |}}  & \ySarah     &  {\tt{\verb|+ 6.2   |}} \\ \hline
   & Total    &  {\tt{\verb|+10.3    |}}  &             &  {\tt{\verb|+42.0   |}} \\
\end{tabular*}
\end{table}

From the last row in Table~\ref{tbl:desresExampleDatabaseState}, $p(\yMary, \yKathy, \ySarah | 9) \allowbreak \approx 1$.  The total information gain in this example, calculated as described in Refs.~\cite{ExperimentalScientificMerit:Knuteson:2007tb,TheoreticalScientificMerit:Knuteson} and approximating $p(\yMary, \yKathy, \ySarah | 9)$ as unity, is
\begin{equation}
{\rm{total\ information\ gain}}  \approx  - \logten{p(\yMary,\yKathy,\ySarah|1)}.
\end{equation}
Unfortunately, $p(\yMary, \yKathy, \ySarah | 1)$ is not in the database.  As indicated by empty space in Table~\ref{tbl:desresExampleDatabaseState}, there is no notion of \yKathy\ or \ySarah\ at time $t=1$.  Use of available information therefore requires assuming the independence of database hypotheses, and for practical purposes using

\begin{eqnarray}
\label{eqn:totalInformationGain}
{\rm{total\ information\ gain}} & \approx & - \left( \logten{{p(\yMary|1)}}  + \right. \nonumber \\
                                &         & \phantom-\phantom(  \ \logten{{p(\yKathy|4)}} +         \nonumber \\
                                &         & \phantom-\phantom(  \left. \logten{{p(\ySarah|8)}}   \right) \nonumber \\
                                & =       & 10.3.
\end{eqnarray}

The evidence provided by each database entry in favor of \yMary, \yKathy, or \ySarah\ is shown in Table~\ref{tbl:desresExampleEvidence}, together with the payoff to the author of each database entry for the publication of that entry.  Assuming a 1\% tax on \$4.2B of revenue in 2020, a total of \$42M is paid to authors of database entries for this product.  Kathy's company gets nearly half (\$20.9M) of this total for making the most surprising discovery.  Dawn makes more (\$8.8M) than Mary does (\$6.2M) by beating her to the punch, although with a less thorough analysis.  In the auction determining the prior of \yKathy, Bidder \#1 makes a fair amount (\$7.1M) at the expense of Bidder \#4 ($-$\$6.3M) for assigning a larger probability that \yKathy\ might be true.  Each separate auction is a zero sum game, with as much money lost as made among auction participants.  Anne loses a bit of change (\$200) for incorrectly providing evidence against \yMary.  Sarah's company makes \$5.5M, somewhat less than they would have made if they hadn't been outguessed in the auction for the prior on \ySarah.  

\section{Previous work}
\label{sec:PreviousWork}

The proposal of this article is the third in a set of proposals of gradually increasing scope.  Reference~\cite{ExperimentalScientificMerit:Knuteson:2007tb} proposes a measure of experimental scientific merit, arguing that more is learned from a surprising experimental result than an unsurprising result, and that the scientific merit of two results should be invariant to whether they are published in a single article or in two separate articles.  The experimental scientific merit of Ref.~\cite{ExperimentalScientificMerit:Knuteson:2007tb} provides a reasonable measure of experimental progress towards answering the big questions in a particular field of science.  It does not address the scientific merit of theoretical work, how the big questions are chosen, or how research funds are allocated among different fields.  

Reference~\cite{TheoreticalScientificMerit:Knuteson} notes that the proposal of Ref.~\cite{ExperimentalScientificMerit:Knuteson:2007tb} applies equally well to theoretical research, can be used to compare experimental and theoretical research on the same footing, and represents an appropriate measure of overall scientific merit.  As in Ref.~\cite{ExperimentalScientificMerit:Knuteson:2007tb}, the proposal in Ref.~\cite{TheoreticalScientificMerit:Knuteson} assumes the existence of the big questions that define a single scientific field, and does not address how the big questions are chosen, or how research funds are allocated among different fields.

The proposal in this article, which extends the proposals in Refs.~\cite{ExperimentalScientificMerit:Knuteson:2007tb,TheoreticalScientificMerit:Knuteson} by providing a market solution to these remaining questions, includes the previous proposals as a limiting case.  Temporarily adopting the notation of these previous articles, let $Y=\{y_1,\ldots,y_j,\ldots,y_m\}$ represent a complete and orthogonal set of distinct states of knowledge as to how nature works, defined with respect to the big questions in some particular field of science, and let $X=\{x_1,\ldots,x_i,\ldots,x_n\}$ represent a complete and orthogonal set of possible outcomes of some experiment.  Assuming $y_j$ is true, let $R(y_j)$ represent the net present value of all revenue that will be generated by products relying on $y_j$ being true, and assume for convenience a tax rate of \onepercent.  The expected payoff to a person who performs the experiment and obtains the result $x_i$ is then
\begin{equation}
\expectedPayoff = \onepercent\ \sum_j{R(y_j) p(y_j|x_i) \logten \frac{p(y_j|x_i)}{p(y_j)}},
\end{equation}
where $p(y_j|x_i)$ is the belief in $y_j$ given the experimental outcome $x_i$, and $p(y_j)$ is the prior on $y_j$.  It may in some cases be plausible to assert that the net present value of all revenue that will be generated by products relying on $y_j$ being true is roughly the same, within considerable estimation uncertainty, for all $j$.  The simplifying assumption $R(y_j)=R$ for all $j$ leads to $\expectedPayoff \propto \informationgain$, where 
\begin{equation}
\informationgain = \sum_j{p(y_j|x_i) \logten \frac{p(y_j|x_i)}{p(y_j)}}
\end{equation}
is the figure of merit proposed in Refs.~\cite{ExperimentalScientificMerit:Knuteson:2007tb,TheoreticalScientificMerit:Knuteson} for evaluating completed scientific research.  If the experiment has not yet been performed, so that the outcome $x_i$ is not yet known, then
\begin{equation}
\expectedPayoff = \onepercent\ \sum_i p(x_i) \sum_j{R(y_j) p(y_j|x_i) \logten \frac{p(y_j|x_i)}{p(y_j)}}.
\end{equation}
The same simplifying assumption that $R(y_j)=R$ for all $j$ leads to $\expectedPayoff \propto \Delta H$, where 
\begin{equation}
\Delta H = \sum_{i,j}{p(x_i,y_j) \logten \frac{p(y_j|x_i)}{p(y_j)}}
\end{equation}
is the expected reduction in information entropy, the figure of merit proposed in Refs.~\cite{ExperimentalScientificMerit:Knuteson:2007tb,TheoreticalScientificMerit:Knuteson} for evaluating proposed scientific research.  From the point of view of this article, Refs.~\cite{ExperimentalScientificMerit:Knuteson:2007tb,TheoreticalScientificMerit:Knuteson} thus implicitly assume equal net present values of revenues generated by products relying on each hypothesis being true, within estimation errors.  ``Qualitatively distinct states of knowledge'' in Refs.~\cite{ExperimentalScientificMerit:Knuteson:2007tb,TheoreticalScientificMerit:Knuteson} are reinterpreted in the context of this proposal to refer to models of nature in which different technologies are possible.

\bibliography{capitalist_science}

\end{document}